\documentclass{article}
\usepackage{ijcai20}

\usepackage{times}

\usepackage{soul}
\usepackage{url}
\usepackage[hidelinks]{hyperref}
\usepackage[utf8]{inputenc}
\usepackage[small]{caption}
\usepackage{graphicx}
\usepackage{amsmath}
\usepackage{booktabs}
\usepackage{color}
\usepackage{enumerate}
\usepackage{algorithm}
\usepackage{algorithmic}
\usepackage{amsfonts}

\usepackage{tabularx}
\usepackage{multirow}

\newcommand{\Dc}{\mathcal{D}}
\newcommand{\Xc}{\mathcal{X}}
\newcommand{\Yc}{\mathcal{Y}}
\newcommand{\Ic}{\mathcal{I}}
\newcommand{\PP}{\mathpzc{P}}
\newcommand{\mS}{\mathrm{S}}
\newcommand{\Z}{\mathbb{Z}}
\newcommand{\mG}{\mathrm{G}}
\newcommand{\mU}{\mathrm{U}}
\newcommand{\xb}{\mathbf{x}}
\newcommand{\R}{\mathbb{R}}
\newcommand{\wb}{\mathbf{w}}

\newcommand{\mL}{\mathrm{L}}
\newcommand{\mcalL}{\mathcal{L}}
\newcommand{\mF}{\mathrm{F}}
\newcommand{\rb}{\mathbf{r}}

\DeclareFontFamily{OT1}{pzc}{}
\DeclareFontShape{OT1}{pzc}{m}{it}{<-> s * [1.200] pzcmi7t}{}
\DeclareMathAlphabet{\mathpzc}{OT1}{pzc}{m}{it}

\newtheorem{definition}{Definition}
\newtheorem{assumption}{Assumption}

\DeclareMathOperator*{\argmax}{argmax}
\DeclareMathOperator*{\argmin}{argmin}

\title{Asymmetrical Vertical Federated Learning}

\author{
Yang Liu$^1$\footnote{Contact Author}\and
Xiong Zhang$^1$\and
Libin Wang$^1$\\
\affiliations
$^1$Tencent Cloud Product Department, Tencent, Shenzhen, China\\
\emails
\{clarkieliu, farleyzhang,benlbwang\}@tencent.com
}

\begin{document}

\maketitle

\begin{abstract}
Federated learning is a distributed machine learning method that aims to preserve the privacy of sample features and labels. In a federated learning system, ID-based sample alignment approaches are usually applied with few efforts made on the protection of ID privacy. In real-life applications, however, the confidentiality of sample IDs, which are the strongest row identifiers, is also drawing much attention from many participants. To relax their privacy concerns about ID privacy, this paper formally proposes the notion of asymmetrical vertical federated learning and illustrates the way to protect sample IDs. The standard private set intersection protocol is adapted to achieve the asymmetrical ID alignment phase in an asymmetrical vertical federated learning system. Correspondingly, a Pohlig-Hellman realization of the adapted protocol is provided. This paper also presents a genuine with dummy approach to achieving asymmetrical federated model training. To illustrate its application, a federated logistic regression algorithm is provided as an example. Experiments are also made for validating the feasibility of this approach.
\end{abstract}

\section{Introduction}

Machine learning algorithms play an essential role in extracting patterns and useful knowledge from datasets. Traditional machine learning methods are usually developed from the centralized or parallel perspective, i.e., the overall datasets are stored in one, or a group of computing machines in hand, and the learning objective can be efficiently achieved by the machines with their direct access to the datasets. Although many of those algorithms are well-studied and have excellent performance over even petabyte-scale datasets, they are still confronted with serious troubles when implemented in the business or research fields with high concerns on data confidentiality. For example, every day portable and wearable smart devices collect billions of users' motion and body condition information, which contains numerous valuable patterns that benefit reproductions. However, it is impossible to perform traditional machine learning algorithms for knowledge extraction, because the distributedly collected data are prohibited to upload to central servers subject to strict privacy laws and regulations, such as the GDPR. Similar examples also exist in a variety of areas, such as finance, health, education, etc.

Federated learning, first proposed in 2016 \cite{mcmahan2016communication}, is a privacy-preserving machine learning approach to analyzing the patterns of distributed sensitive datasets. The notion was later expanded to a clearer and more comprehensive framework with three components: horizontal federated learning in which participants share the same feature space, vertical federated learning in which participants share the same ID space, and federated transfer learning for disjoint data distributions. In this framework, every distributed participant shares encrypted messages that are computed based on their individual datasets and updates its local model according to the received messages, with or without the presence of a trusted third party. However, in federated learning, especially in the vertical scenario, it is common but unreasonable to omit the investigation of the sample ID's information leakage after leveraging them to align distributed datasets. In a standard vertical federated learning system, for example, samples held by different participants are aligned by executing some secure protocol and letting everyone know what the exact intersect set is. In practical applications, the participants are usually companies or institutions in competitive relations, and many clients of one company are potential advertising targets of another. On the one hand, thus, the participants would not allow the disclosure of their sample IDs. On the other hand, there are asymmetrical federations in real life, in which a subset of the participants are small companies with strong requirements of ID privacy protection, while the other are large companies who do not concern much ID privacy, because their guests are almost all citizens in the society. Such an unbalanced setup requires a vertical federated learning system to distinguish the ``weak" and the ``strong" sides of the federation and take into consideration their specific privacy protection demands.

In this paper, we specify the vertical federated learning into two classes: the symmetrical and the asymmetrical, to develop federated learning algorithms that preserve the privacy of sample IDs for the participants who indeed demand. The contributions of this paper are summarized as follows.
\begin{enumerate}[(i)]
	\item We formally propose and comprehensively characterize the notion of asymmetrical vertical federated learning.
	\item We incorporate the standard private set intersection protocol to achieve the asymmetrical ID alignment phase in an asymmetrical vertical federated learning system. In addition, we provide a Pohlig-Hellman realization of the adapted private set intersection protocol.
	\item We present a genuine with dummy approach to achieving asymmetrical federated model training. To illustrate its application, we provide a federated logistic regression algorithm as an example. Experiments are also made for validating the feasibility of the approach.
\end{enumerate}

The rest of the paper is organized as follows. In Section \ref{sec:prob_def}, we formulate the symmetrical and asymmetrical classification of vertical federated learning. We further present an asymmetrical private set intersection protocol for ID alignment and one of its realizations in Section \ref{sec:id_alignment}. Section \ref{sec:model_training} provides the genuine with dummy approach to asymmetrical model training and its application to federated logistic regression, along with a few experimental validations in Section \ref{sec:experiments}. Finally, several concluding remarks are given in Section \ref{sec:conclusions}.

\section{Problem Definition}\label{sec:prob_def}

In this section, we revisit the notion of vertical federated learning and formally categorize it into two classes to study the privacy preservation of sample IDs.

\subsection{Vertical Federated Learning}

\begin{figure}[ht]
	\centering
	\includegraphics[width=9.2cm]{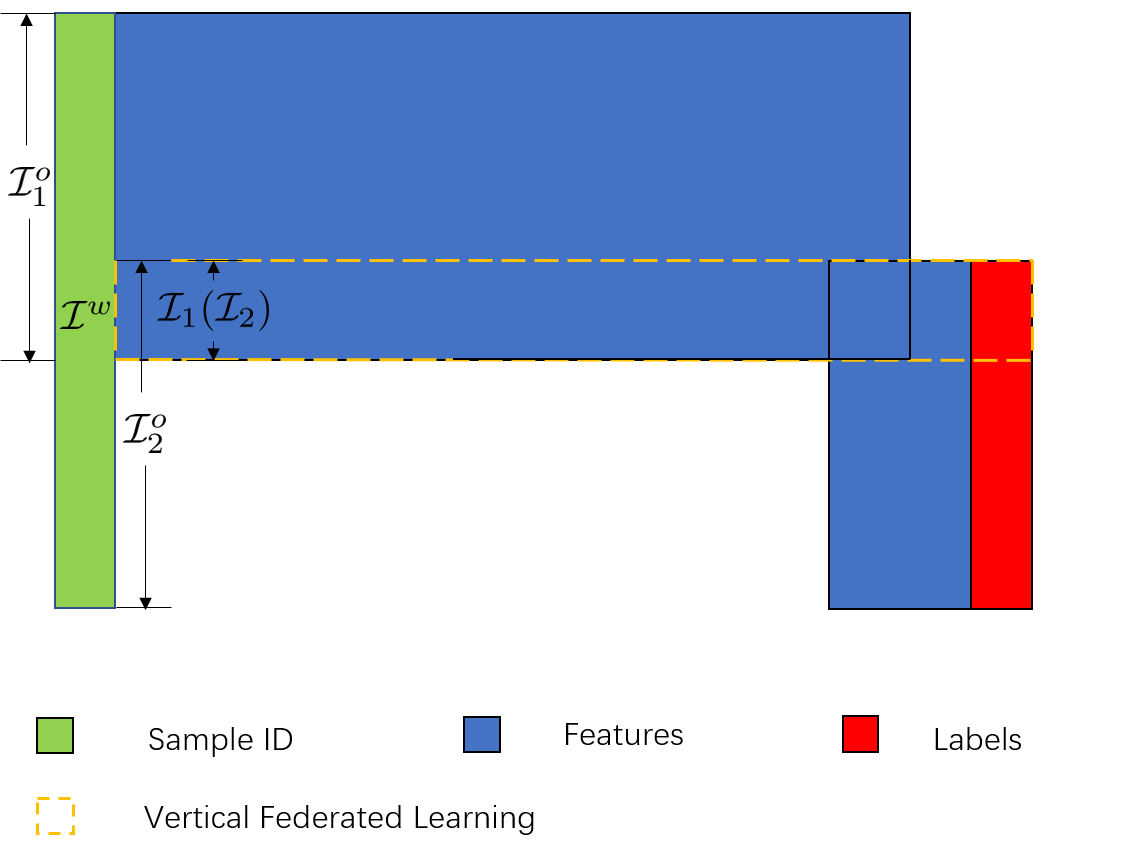}
	\caption{Symmetrical Vertical Federated Learning}
	\label{fig:vertical}
\end{figure}

Let $\Dc=(\Ic,\Xc,\Yc)$ denote a complete dataset with $\Ic,\Xc,\Yc$ representing the sample ID space, the feature space and the label space, respectively.  It was defined in \cite{yang2019federated} that the vertical federated learning is conducted over two datasets $\Dc_1=(\Ic_1,\Xc_1,\Yc_1)$, $\Dc_2=(\Ic_2,\Xc_2,\Yc_2)$ satisfying
$$ \Xc_1\neq\Xc_2, \Yc_1\neq\Yc_2,\Ic_1=\Ic_2.$$
In real-world applications, however, it is nearly impossible to find two original datasets collected by distributed parties that share exactly the same sample ID space. Therefore, as a preparation phase for vertical federated learning, it is necessary to introduce appropriate ID-alignment protocols that assist each party with its secure identification of $\Ic_1=\Ic_2$ and establishment of two datasets' row mapping.  We define $\Dc_1^o=(\Ic_1^o,\Xc_1,\Yc_1), \Dc_2^o=(\Ic_2^o,\Xc_2,\Yc_2)$ as the pre-ID-alignment datasets that correspond to $\Dc_1$ and $\Dc_2$, respectively. Clearly, $\Ic_1\subset\Ic_1^o$ and $\Ic_2\subset\Ic_2^o$. In addition, we write $\Ic^w$ as the whole ID space that contains all possible elements in $\Dc_1^o$ and $\Dc_2^o$. Then the vertical federated learning presented by \cite{yang2019federated}  is depicted in Figure \ref{fig:vertical}.

\subsection{Private Set Intersection}

Recall that ID-alignment protocols are essential in the preparation of  vertical federal model training. To achieve ID alignment, Private Set Intersection (PSI) protocols, as one of the most well studied areas in secure multiparty computation, are usually implemented in a federated learning system. In standard PSI, each party $\PP_i$ holds a set $\mS_i$, which involves $\PP_i$'s confidential data. All parties would like to cooperatively find the intersection $\mS^\prime=\bigcap_i\mS_i$, and in the meantime, each $\PP_i$ keeps the elements in $\mS_i\setminus\mS^\prime$ private. The realization of PSI protocols can be based on classical public-key cryptosystems \cite{freedman2004efficient,liang2004privacy}, oblivious transfer \cite{pinkas2014faster},  garbled circuits \cite{huang2012private}, etc.

\subsection{Symmetrical and Asymmetrical}

Let us call the parties $\PP_1, \PP_2$ who own $\Dc_1^o, \Dc_2^o$. As can be seen from Figure \ref{fig:vertical}, $\PP_1$ and $\PP_2$ have equally powerful positions in the sense that the numbers of elements in their pre-ID-alignment sample ID spaces $\Ic_1^o,\Ic_2^o$ share the same order of magnitude and both sets play a major role in the whole ID space $\Ic^w$.
In fact, even if $\PP_1$ picks $e\in\Ic^w$ uniformly at random, there is $\left|\Ic_2^o\right|/\left|\Ic^w\right|\gg0$ probability of $e\in\Ic_2^o$. In such ``federation of the strong", each party would not gain much knowledge by executing PSI protocols to obtain the  intersection $\Ic_1=\Ic_2$, and thus it bothers neither party to let the other party know what IDs it possesses. Since finally both parties symmetrically obtain the intersection for model training, we term this scenario symmetrical vertical federated learning.

\begin{figure}[ht]
	\centering
	\includegraphics[width=9.2cm]{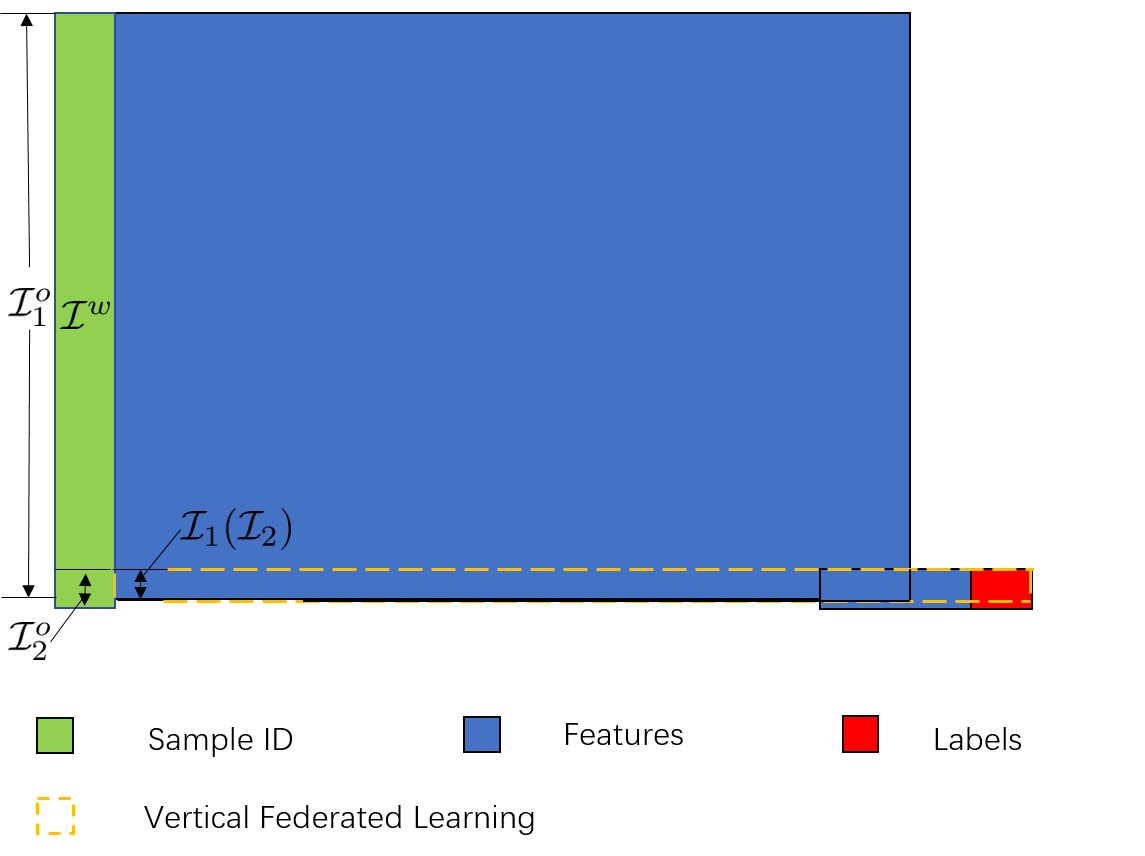}
	\caption{Asymmetrical Vertical Federated Learning}
	\label{fig:asym}
\end{figure}

The opposite scenario is that $\PP_2$, without loss of generality, has $\left|\Ic_2^o\right|/\left|\Ic^w\right|\approx0$, while $\PP_1$ still holds $\left|\Ic_1^o\right|/\left|\Ic^w\right|\gg0$. It is common that the federated learning system to be established is for $\PP_2$'s learning task, and thus the labels are provided by $\PP_2$. This vertical data distribution is as shown in Figure \ref{fig:asym}. Clearly,  there is a sufficiently small probability that an arbitrary ID $e\in\Ic^w$ one picks uniformly at random exactly belongs to $\Ic_2^o$. Therefore, we say $\PP_2$ is at the weak side in the sense that each sample ID in $\Ic_2^o$ is regarded as sensitive information, whose privacy would be severely compromised through the revelation of $\Ic_1=\Ic_2$ by executing standard PSI protocols. In such ``federation of the weak and the strong", it is necessary to asymmetrically protect the ID privacy of the weak party in the ID alignment phase. This scenario is termed asymmetrical vertical federated learning.

It is reasonable not to analyze the ``federation of the weak" scenario, because we may shrink the whole space $\Ic^w$ to $\Ic^w\setminus(\Ic_1^o\cup\Ic_2^o)$ so that this scenario can be reduced to the ``federation of the strong" one. Throughout this paper, we impose the following assumption.

\begin{assumption}\label{ass:iw_union}
	$\Ic^w = \Ic_1^o \cup \Ic_2^o$.
\end{assumption}

It is evident that if Assumption \ref{ass:iw_union} holds, either $\left|\Ic_1^o\right| \ge \frac{1}{2}\left|\Ic^w\right|$ or $\left|\Ic_2^o\right| \ge \frac{1}{2}\left|\Ic^w\right|$. Now by representing a positive number $n$ by $n=a\cdot 10^b$ with $a\in[10^{-\frac{1}{2}},10^{\frac{1}{2}})$ and naming $b$ the order of magnitude of $n$, we provide the following precise definition for.

\begin{definition}\label{def:svfl_avfl}
	Under Assumption \ref{ass:iw_union}, vertical federated machine learning can be classified into two categories based on the participating parties' pre-ID-alignment sample ID spaces.
	\begin{enumerate}[(i)]
		\item It is called Symmetrically Vertical Federated Learning (SVFL) if $$-\frac{1}{2}\le\log_{10}\frac{\left|\Ic_1^o\right|}{\left|\Ic^w\right|},\log_{10}\frac{\left|\Ic_1^o\right|}{\left|\Ic^w\right|} \le 0 .$$
		\item It is called Asymmetrical Vertical Federated Learning (AVFL) if either one of the following two inequalities holds.
		\begin{align}
		&\log_{10}\frac{\left|\Ic_1^o\right|}{\left|\Ic^w\right|} < -\frac{1}{2},\label{eq:avfl_con1}\\
		&\log_{10}\frac{\left|\Ic_2^o\right|}{\left|\Ic^w\right|} < -\frac{1}{2}.\label{eq:avfl_con2}
		\end{align}
		In AVFL, we further name $\PP_1$ (or $\PP_2$) the weak participant if (\ref{eq:avfl_con1}) (or (\ref{eq:avfl_con2})) holds, and $\PP_2$ (or $\PP_1$) the strong participant.
	\end{enumerate}
\end{definition}
Since Assumption \ref{ass:iw_union} holds in Definition \ref{def:svfl_avfl}, there always exist a weak and a strong participant in AVFL. 
Note that we base the analysis above and the rest of the paper on the two-party federation case for the simplicity of demonstration. In fact, the analysis can be trivially extended to the multi-party case by asymmetrically protecting the sample ID privacy of the weak party.

\section{Asymmetrical ID Alignment}\label{sec:id_alignment}

ID alignment is typically the first stage of a vertical learning workflow. In this section, we adapt the standard PSI protocol to achieve asymmetrical ID alignment and provide a realization using a classical cryptosystem.

\subsection{Asymmetrical PSI Protocols}

Recall that in SVFL's ID alignment phase, the execution of standard PSI protocols not only lets each party gain knowledge about the samples they both hold, but also provides a tag that manages to efficiently link together distributed sample fragments and paves the way for the follow-up federated model training phase. To realize a proper ID-alignment phase in AVFL, one has to adapt standard PSI protocols so that they satisfy
\begin{enumerate}[(i)]
	\item The exact intersection set $\Ic_1=\Ic_2$ is kept private against the strong participant. 
	\item In the federated model training phase, the distributed samples with ID in $\Ic_1=\Ic_2$ are still alignable.
\end{enumerate}
Let $\Ic_w^o,\Ic_s^o$ denote the pre-ID-alignment sample ID sets held by the weak and the strong participant, respectively. Alternatively, they are defined by
\begin{align*}
\{\Ic_w^o\}=\argmin_{\Ic\in\{\Ic_1^o,\Ic_2^o\}}\left|\Ic\right|,\ \{\Ic_s^o\}=\argmax_{\Ic\in\{\Ic_1^o,\Ic_2^o\}}\left|\Ic\right|.
\end{align*}
A possible ID-alignment approach that meets requirements (i)-(ii) above is to use a variant of PSI protocols that we present below.

\begin{figure}[ht]
	\centering
	\includegraphics[width=9.2cm]{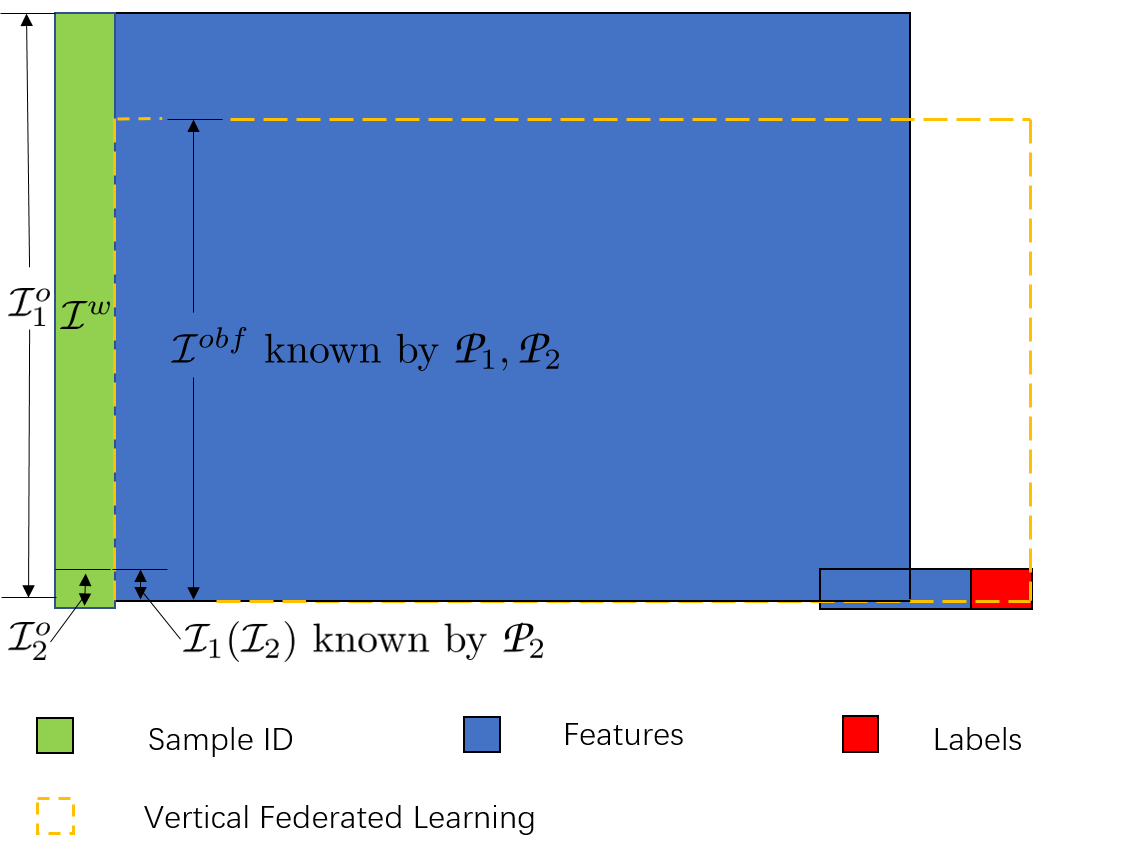}
	\caption{Asymmetrical PSI Protocols for AVFL}
	\label{fig:asym_intersection}
\end{figure}

\begin{definition}\label{def:apsi}
	Asymmetrical PSI (APSI) protocols, as a variant of PSI protocols, yield an obfuscated set $\Ic^{obf}\subset \Ic^w$ at each party satisfying
	$$ \Ic_1=\Ic_2\subset\Ic^{obf}\subset\Ic_s^o.$$
	In addition, only the weak participant further knows $\Ic_1=\Ic_2$. 
\end{definition}
 The output of such PSI protocols is illustrated in Figure \ref{fig:asym_intersection}.

\subsection{A Pohlig-Hellman Realization}

We now provide a realization of APSI protocols based on the Pohlig-Hellman encryption. Define $\Z_n=\{0,1,\dots,n-1\}$ and $\Z_n^\ast\subset\Z_n$ as the set of integers coprime to $n$. Let $\mG_n=(\Z_n^\ast,\cdot)$ denote a multiplicative group. Then the well-known Pohlig-Hellman encryption scheme is described by the following three components.
\begin{enumerate}[(i)]
	\item (Key generation) Select a prime number $p$ such that every plaintext is an element of $\mG_p$, where $p-1$ has at least one large prime factors. For example, select $p=2q+1$ with $q$ being also a prime number. Then select $a\in\mG_{p-1}$ and compute $a^{-1}\in\mG_{p-1}$. Finally, one reveals $\mG_p$ as public knowledge and keeps $(a,a^{-1})$ as the key to this symmetrical cryptosystem.
	\item (Encryption) Encrypt the plaintext $m\in\mG_p$ by
	$$ E_{a}(m) = m^a. $$
	\item (Decryption) Decrypt the ciphertext $c\in\mG_p$ by
	$$ D_{a}(c) = c^{a^{-1}}. $$
\end{enumerate}
It is straightforward to show the commutative property of the Pohlig-Hellman encryption by checking $E_a\circ E_b = E_b\circ E_a$ for any $a,b\in\mG_{p-1}$. Now we present an APSI protocol based on the Pohlig-Hellman encryption in Algorithm \ref{alg:apsi}.

\begin{algorithm}[ht]
	Input: The strong participant $\PP_1$ holds $\Ic_1^o$. The weak participant $\PP_2$ holds $\Ic_2^o$. Security number $\lambda\in[0,1]$.\\
	Output: $\PP_1$ only obtains $\Ic^{obf}$ satisfying $\Ic_1^o\cap\Ic_2^o\subset\Ic^{obf}$ and $$\frac{\left|\Ic^{obf}\right|}{\left|\Ic_1^o\cap\Ic_2^o\right|}=h(\lambda)\overset{\triangle}{=}\bigg(\frac{\left|\Ic_1^o\right|}{\left|\Ic_1^o\cap\Ic_2^o\right|}\bigg)^\lambda.$$
	$\PP_2$ obtains both $\Ic_1^o\cap\Ic_2^o$ and $\Ic^{obf}$.
	\begin{algorithmic}[1]
		\STATE $\PP_1$ and $\PP_2$ negotiate a public group $\mG_p$ for Pohlig-Hellman encryption.
		\STATE $\PP_1$ and $\PP_2$ privately generates keys $(a,a^{-1}),(b,b^{-1})\in\mG_{p-1}\times\mG_{p-1}$, respectively.
		\STATE $\PP_1$ and $\PP_2$ computes $\mU_1=\{E_a(x):x\in\Ic_1^o\}$ and $\mU_2=\{E_b(x):x\in\Ic_2^o\}$, respectively.
		\STATE $\PP_1$ and $\PP_2$ exchange $\mU_1,\mU_2$.
		\STATE $\PP_1$ and $\PP_2$ computes $\mU_2^\dagger=\{E_a(x):x\in\mU_2\}$ and $\mU_1^\dagger=\{E_b(x):x\in\mU_1\}$, respectively.
		\STATE $\PP_1$ sends $\mU_2^\dagger$ to $\PP_2$.
		\STATE $\PP_2$ randomly selects $\mU^{obf\dagger}$ satisfying $\mU_1^\dagger\cap\mU_2^\dagger\subset\mU^{obf\dagger}\subset\mU_1^\dagger$ and 
		$\left|\mU^{obf}\right|=h(\lambda)\left|\mU_1^\dagger\cap\mU_2^\dagger\right|.$
		\STATE 	$\PP_2$ sends $\mU^{obf\dagger}$ to $\PP_1$.
		\STATE $\PP_1$ computes $\mU^{obf}=\{D_a(x):x\in\mU^{obf\dagger}\}$.
		\STATE $\PP_1$ sends $\mU^{obf}$ to $\PP_2$.
		\STATE $\PP_2$ computes $\Ic^{obf}=\{D_b(x):x\in\mU^{obf}\}$ and $ \Ic_1^o\cap\Ic_2^o=\Ic_2^o\cap\Ic^{obf}$.
		\STATE $\PP_2$ shares $\Ic^{obf}$ to $\PP_1$.
	\end{algorithmic}
	\caption{Pohlig-Hellman APSI Protocol}
	\label{alg:apsi}
\end{algorithm}

\medskip

\noindent{\em Security Analysis.} In Algorithm \ref{alg:apsi}, the exchanged information by step 6 is $\mU_1,\mU_2,\mU_2^\dagger$, whose security is evidently guaranteed by the Pohlig-Hellman encryption scheme. By step 8, the strong participant $\PP_1$ receives the encrypted obfuscated set $\mU^{obf\dagger}$ but it cannot distinguish which elements belong to the encrypted true intersection set. In fact, $\PP_1$ cannot even obtain any plaintext based on $\mU^{obf\dagger}$ because the elements in $\mU^{obf\dagger}$ are still encrypted with $\PP_2$'s key $b$. By step 11, since the cooperatively decrypted message is only $\Ic^{obf}$, $\PP_1$'s private information $\Ic_1^o\setminus\Ic^{obf}$ is protected against the weak participant $\PP_2$.  Clearly, in step 12, only $\Ic^{obf}$ is revealed to $\PP_1$ with $\Ic_1^o\cap\Ic_2^o$ kept private by $\PP_2$.

It can be noted that the security number $\lambda$ is interpreted as a cardinality ratio $h(\lambda)$ of the obfuscated intersection set $\Ic^{obf}$ to the true intersection set $\Ic_1=\Ic_2$. As can be directly computed, there is $1/h(\lambda)$ probability that a uniformly random element in $\Ic^{obf}$ that $\PP_1$ picks belongs to $\Ic_1=\Ic_2$. Clearly, $h(0)=1$, in which case $\Ic^{obf}=\Ic_1=\Ic_2$ and $\PP_1,\PP_2$ both obtain the true intersect set. In this special case, AVFL becomes SVFL. As $\lambda$ goes up from zero to one, $1/h(\lambda)$ exponentially decreases, and thereby it becomes more difficult for $\PP_1$ to potentially identify $\Ic_1=\Ic_2$ from $\Ic^{obf}$. When $\lambda$ reaches the maximum one,  the obfuscated set $\Ic^{obf}=\Ic_1^o$, i.e., the whole ID space of $\PP_1$ is used for obfuscating and $\PP_1$ cannot gain any knowledge by executing Algorithm \ref{alg:apsi}.

\section{Asymmetrical Federated Model Training}\label{sec:model_training}

In this section, we investigate the asymmetrical federated model training process and propose a novel and general approach to training a model in an asymmetrical fashion that is as good as in a symmetrical fashion. We also provide an application of this approach to an existing federated learning algorithm.

\subsection{Genuine with Dummy Approach}
Using APSI protocols given in Definition \ref{def:apsi}, the weak participant obtains the true intersection set $\Ic_1=\Ic_2$, which is a subset of the obfuscated intersection set $\Ic^{obf}$ that the strong participant knows. As shown in Figure \ref{fig:asym_intersection}, the vertical federated learning domain now contains margins, i.e., all labels and a few features are missing for the samples with their ID in $\Ic^{obf}\setminus\Ic_1$. Indeed, federated transfer learning \cite{yang2019federated} can be used to fill in the margins by the feature-representation-transfer approach \cite{pan2009survey}. Nevertheless, in many practical applications, such as financial risk management, it is normal to train machine learning model based on the small but exactly original data in $\Ic_1=\Ic_2$, in order to avoid misjudgment of dishonest conduct. In these areas, the learned features and labels for $\Ic_{obf}\setminus\Ic_1$, even if not ``negatively transferred", may lead to undesirably strict or loose risk control strategies. Therefore, it is necessary to design asymmetrical model training schemes that take the distributed output of APSI protocols and yield the same or almost the same result as the SVFL. Based on the standard vertical model training in \cite{yang2019federated}, we now present a Genuine with Dummy (GWD) approach to achieving asymmetrical model training as follows. Note that a trusted third party $\PP_3$ is introduced as a secure coordinator.

\begin{enumerate}[(i)]
	\item $\PP_3$ generates a public-key cryptosystem, and sends the public key to $\PP_1,\PP_2$.
	\item $\PP_1,\PP_2$ exchange intermediate variables to cooperatively compute gradient and loss. The weak participants, say $\PP_2$, normally execute computation rule for the samples in $\Ic_1$, i.e., the genuine, but set the variables that correspond to the samples in $\Ic^{obf}\setminus\Ic_1$, i.e., the dummy, to specific mathematical identities so that their existence will not affect the relevant computed result. The identities can be, for example, zero in addition, one in multiplication or $f(x)=x$ in function composition.
	\item $\PP_1,\PP_2$ turn to $\PP_3$ for gradient and loss decryption service, and update their local models.
\end{enumerate}

As we can see, the central idea of the GWD approach provided above is to let the weak participant execute the normal protocol for the genuine samples, while mathematically mute the dummy samples that it actually does not hold before sending them to the strong participant. To keep the strong participant unaware of the existence of the dummy samples, a potential method is to implement semantically secure encryption scheme, such as Paillier cryptosystem \cite{paillier1999public}, to disable the participant from efficiently distinguishing the identities out of a group of normal variables. In addition, it is clear that the asymmetrical model training achieved with the GWD approach would exhibit exactly the same performance as the standard (or symmetrical) model training because the introduced identities strictly guarantee invariant intermediate results at every step.

\begin{figure}[ht]
	\centering
	\includegraphics[width=8.5cm]{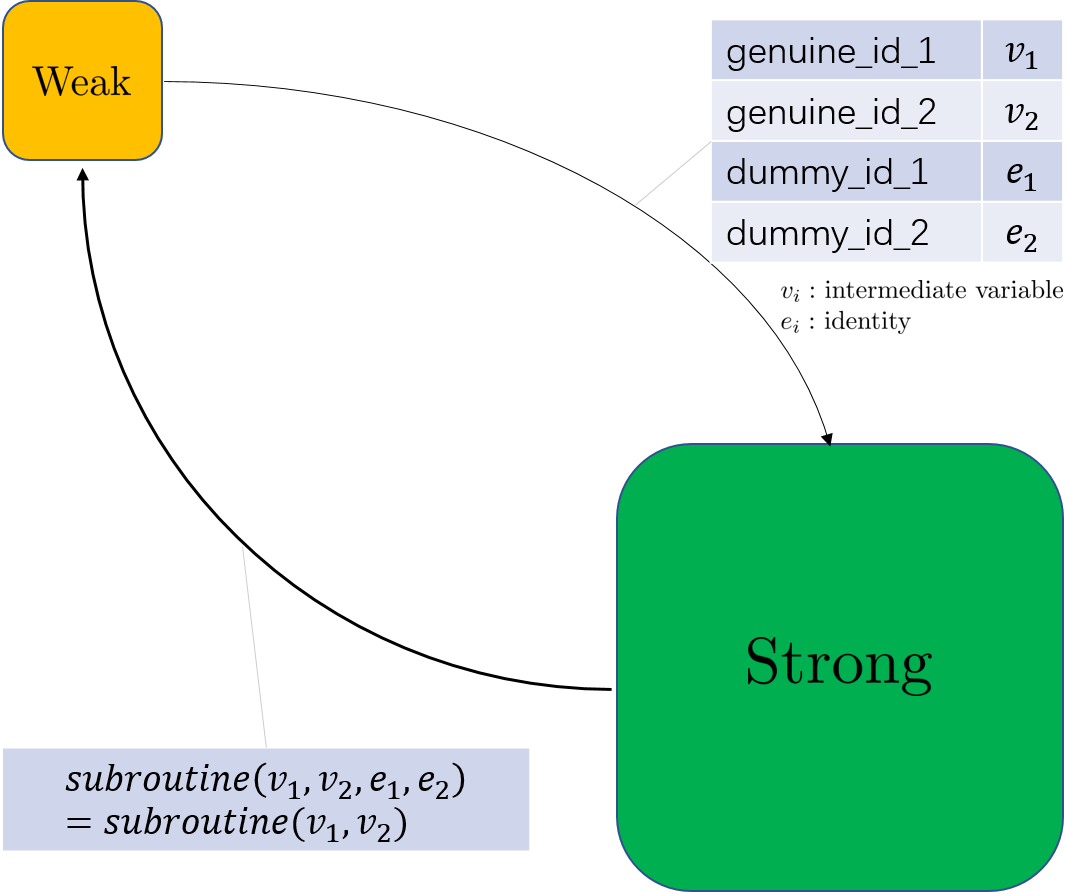}
	\caption{Architecture of the GWD approach}
	\label{fig:gwd}
\end{figure}

An illustration of GWD architecture is provided in Figure \ref{fig:gwd}, which depicts the federated execution process of a general subroutine. The weak participant would like to send encrypted messages $v_1,v_2$ to the strong participant and then expect a response of the execution result $subroutine(v_1,v_2)$. However, the direct transmission would give away the genuine IDs. Instead,  the weak particpant sends $v_1,v_2$ along with the mathematical identities $e_1,e_2$ corresponding to the dummy samples, and finally receives $subroutine(v_1,v_2,e_1,e_2)$, which is identical to what it expects. Besides, the strong participant performs the computation $subroutine(\cdot)$, while remaining oblivious to the genuine IDs.

\subsection{Asymmetrical Vertical Logistic Regression}

We now take the coordinator-free federated logistic regression training presented in \cite{yang2019parallel} as an example and adapt it to an asymmetrical training protocol using the GWD approach. Let $\langle M \rangle$ denote the ciphertext of a plain message $M$. Then the adapted Asymmetrical Vertical Logistic Regression (AVLR) training protocol is presented in Algorithm \ref{alg:asym_lr}.

\begin{algorithm}
	Input: The strong participant $\PP_1$ holds the sample set $(e_i,\xb_i^s)\in\Ic^{obf}\times\R^{m^s}$. The weak participant $\PP_2$ holds the sample set $(e_i,\xb_i^w,y_i^w)\in\Ic_1^o\cap\Ic_2^o\times\R^{m^w}\times\{0,1\}$. Learning rate $\eta>0$.\\
	Output:  $\PP_1,\PP_2$ learn the weights $\wb_\ast^s\in\R^{m^s},\wb_\ast^w\in\R^{m^w}$, respectively, such that the joint weight $\wb_\ast=(\wb_\ast^{s\top},\wb_\ast^{w\top})^\top$ is the global optimum of the model.
	\begin{algorithmic}[1]
		\STATE $\PP_2$ instantiates a semantically secure additively homomorphic cryptosystem $crypto$.
		\STATE $\PP_1$ and $\PP_2$ initialize their model estimates $\wb_0^s\in\R^{m^s}$ and $\wb_0^w\in\R^{m^w}$, respectively.
		\WHILE {$k=0,1,2,\dots$}
		\STATE $\PP_1$ sends $\mL_k^{obf}=\{(e_i,\wb_k^{s\top}\xb_i^s):e_i\in\Ic^{obf}\}$ to $\PP_2$.
		\STATE Based on $\mL_k^{obf}$, $\PP_2$ computes $l_{ik}=\wb_k^{s\top}\xb_i^s + \wb_k^{w\top}\xb_i^w$ for all $e_i\in\Ic_1^o\cap\Ic_2^o$, and further
		\begin{equation}\notag
		\phi_{ik} = \left\{
		\begin{aligned}
		&y_i - \big(1+\exp(-l_{ik})\big)^{-1}\textnormal{ if } e_i\in\Ic_1^o\cap\Ic_2^o;\\
		&0\qquad\qquad\qquad\textnormal{ if }e_i\in\Ic^{obf}\setminus(\Ic_1^o\cap\Ic_2^o).
		\end{aligned}
		\right.
		\end{equation}
		\STATE $\PP_2$ uses $crypto$ to encrypt $\phi_{ik}$ for all $i\in\Ic^{obf}$ and sends $\mF_k^{obf}=\{(e_i,\langle \phi_{ik} \rangle):e_i\in\Ic^{obf}\}$ to $\PP_1$.
		\STATE $\PP_2$ computes the global likelihood and its local gradient in plaintext space
		\begin{align*}
		\mcalL(\wb_k) &= \frac{1}{\left|\Ic_1^o\cap\Ic_2^o\right|} \sum\limits_{e_i\in\Ic_1^o\cap\Ic_2^o} y_i l_{ik} \\
		&\quad- \log(1+\exp(l_{ik})),\\
		\nabla^w \mcalL(\wb_k) &= \frac{1}{\left|\Ic_1^o\cap\Ic_2^o\right|} \sum\limits_{e_i\in\Ic_1^o\cap\Ic_2^o} \phi_{ik}\xb_i^w.
		\end{align*}
		\STATE Based on $\mF_k^{obf}$, $\PP_1$ computes its local gradient in cipherspace
		\begin{align*}
		\langle \left|\Ic_1^o\cap\Ic_2^o\right|\cdot\nabla^s \mcalL(\wb_k) \rangle = \sum\limits_{e_i\in\Ic^{obf}} \langle \phi_{ik} \rangle \xb_i^s,
		\end{align*}
		\STATE $\PP_1$ further masks its local gradient by $G^s = \rb_k \odot \langle \left|\Ic_1^o\cap\Ic_2^o\right|\cdot\nabla^s \mcalL(\wb_k) \rangle$, where $\rb\in\R^{m^s}$ is randomly chosen and $\odot$ denotes the Hadamard product. $\PP_1$ further sends $G^s$ to $\PP_2$ for decryption.
		\STATE $\PP_2$ uses $crypto$ to decrypt $G^s$ and divides the result by $\left|\Ic_1^o\cap\Ic_2^o\right|$. Finally, $\PP_2$ manages to send back $\rb_k\odot\nabla^s \mcalL(\wb_k)$ to $\PP_1$.
		\STATE $\PP_1$ uses the Hadamard division to recover $\nabla^s \mcalL(\wb_k)$.
		\STATE $\PP_1$ and $\PP_2$ update their local models by
		\begin{align*}
		\wb_{k+1}^s &= \wb_k^s + \eta \nabla^s\mcalL(\wb_k),\\
		\wb_{k+1}^w &= \wb_k^w + \eta \nabla^w\mcalL(\wb_k),
		\end{align*}
		respectively.
		\ENDWHILE
	\end{algorithmic}
	\caption{Asymmetrical Vertical Logistic Regression Model Training}
	\label{alg:asym_lr}
\end{algorithm}

In Algorithm \ref{alg:asym_lr}, the central subroutine is to let the weak participant $\PP_2$ share the encrypted scalars $\langle \phi_{ik} \rangle$, based on which the strong participant $\PP_1$ evaluates its local gradient $\langle \nabla^s \mcalL(\wb_k) \rangle$. To implement the GWD approach, $\PP_2$ computes $\phi_{ik}$ for $e_i\in\Ic_1^o\cap\Ic_2^o$ based on the true labels, but set $\phi_{ik}=0$ for all $e_i\in\Ic^{obf}\setminus(\Ic_1^o\cap\Ic_2^o)$, which is the addition identity. Due to the fact that $crypto$ preserves semantic security, $\PP_2$ can neither identify $\langle 0 \rangle$ from the received set of $\phi_{ik}$s, nor distinguish which samples are the dummies. Since $\PP_1$ performs additions to compute $\nabla^s \mcalL(\wb_k)$, the presence of the dummy samples' $\phi_{ik}$ would have no effects on the result. In this way, the dataset, over which the federation of $\PP_1$ and $\PP_2$ train their model, is inherently the sample set with IDs in $\Ic_1^o\cap\Ic_2^o$. This guarantees that the asymmetrical vertical training yields the same result as the standard (symmetrical) vertical training does.

\section{Experiments}\label{sec:experiments}

This section provides a few experiments that validate the feasibility of the APSI + AVLR protocol and compare it with the existing standard (symmetrical) protocol.

\subsection{Settings}

We implement our APSI and AVLR algorithms in the federated learning framework FATE\footnote{\url{https://github.com/FederatedAI/FATE}}. The performance of the APSI + AVLR protocol is demonstrated over the dataset MNIST\footnote{\url{http://yann.lecun.com/exdb/mnist/}}, which has 60000 samples and 784 features. To adapt the dataset to our distributed setup, we manually allocate an ID to each sample, split and assign the partitions to the weak participant $\PP_2$ and the strong participant $\PP_1$ as in Table \ref{tab:mnist}.

\begin{table}[htbp]
	\centering
	\begin{tabularx}{0.5\textwidth}{| X | X | X |}
		\hline
		     & 392 Features     & 392 Features     \\ \hline
		10000 Samples         & $\PP_2$     & \multirow{2}{\hsize}{$\PP_1$}         \\ \cline{1-2}
		50000 Samples         & (Abandoned)         &          \\ \hline
	\end{tabularx}
	\caption{MNIST Dataset Partitioning}
	\label{tab:mnist}
\end{table}

As indicated by Table \ref{tab:mnist}, the underlying federated training would be performed over the $10000$ samples that both $\PP_1$ and $\PP_2$ hold. Therefore, the training performance is expected not to be as good as other algorithms that take the whole dataset as input. However, it is fairly reasonable to utilize the distributed setup in Table \ref{tab:mnist}, since the experiments are conducted for the purpose of validating the feasibility of the APSI + AVLR protocol and comparing it with the standard (symmetrical) version.

For the computing hardware, we use two individual machines to serve as the participants and either of them has 4 CPU cores and 16GB RAM. These machines are both located in the same region of Tencent Cloud\footnote{\url{https://cloud.tencent.com/}}.

\subsection{Numerical Results}

\begin{figure}[ht]
	\centering
	\includegraphics[width=9.4cm]{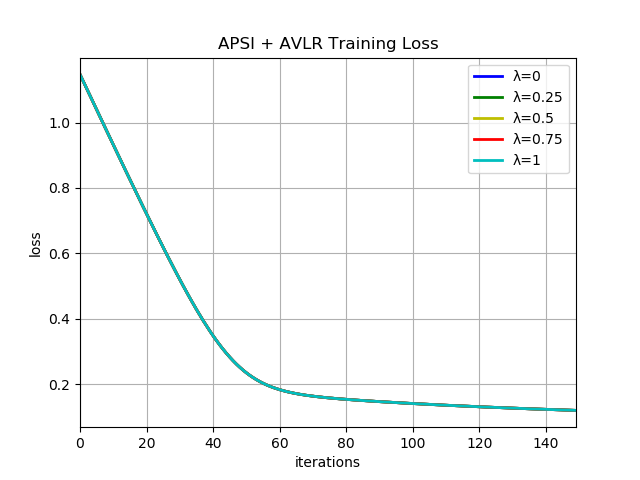}
	\caption{Training Loss Trajectories of the APSI + AVLR Protocol}
	\label{fig:loss}
\end{figure}

\begin{figure}[ht]
	\centering
	\includegraphics[width=9.4cm]{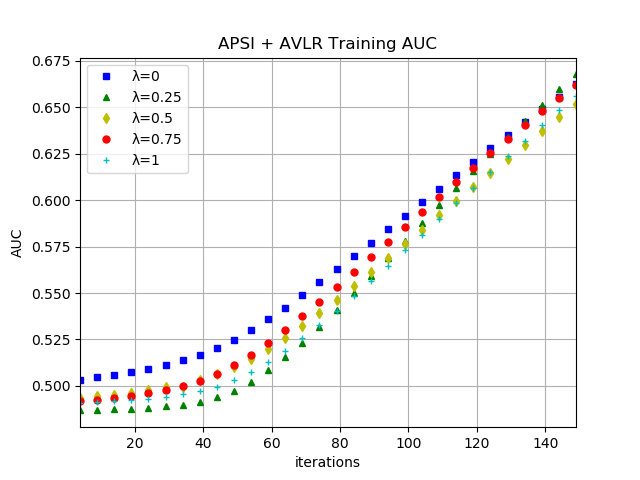}
	\caption{Training AUC Trajectories of the APSI + AVLR Protocol}
	\label{fig:auc}
\end{figure}

In the experiments of the APSI + AVLR protocol, we adopt the fixed learning rate $\eta=0.15$ but various security numbers $\lambda=0,0.25,0.5,0.75,1$, and let the training process execute for $150$ iterations. It is worth mentioning the standard (symmetrical) model training corresponds the $\lambda=0$ case in our experiments. We plot the trajectories of training loss and AUC in Figure \ref{fig:loss} and Figure \ref{fig:auc}, respectively. In Figure \ref{fig:loss}, the trajectories are almost the same for different $\lambda$s, especially for the $\lambda=0$ and the $\lambda\neq0$ case, which also holds true for the AUC trend in Figure \ref{fig:auc}. The observation of these figures validates that the APSI + AVLR protocol has as good performance as its symmetrical version due to the introduction of mathematical identities.

\section{Conclusions}\label{sec:conclusions}

In this paper, we studied the privacy preservation of sample IDs in vertical federated learning. To meet the privacy protection demands of different participants, we first proposed the notion of asymmetrical vertical federated learning. We then adapted the standard private set intersection protocol to achieve the asymmetrical ID alignment phase in an asymmetrical vertical federated learning system. Correspondingly, a Pohlig-Hellman realization of the adapted protocol was provided. To achieve asymmetrical federated model training, we also presented a genuine with dummy approach. We illustrated its application by providing a federated logistic regression as an example. Experiments were also made for validating the feasibility of this approach.

\bibliographystyle{named}
\bibliography{asym_fed_learning}

\end{document}